\documentclass[onecolumn,preprintnumbers,amsmath,amssymb,superscriptaddress]{revtex4}
\usepackage{graphicx}
\usepackage{dcolumn}
\usepackage{bm}

\begin{document}



\title{Limits on the WIMP-nucleon scattering cross-section from neutrino telescopes}

\author{G. Wikstr\"{o}m}
\thanks{E-mail address: wikstrom@fysik.su.se}
\author{J. Edsj\"o}
\thanks{E-mail address: edsjo@fysik.su.se}
\affiliation{The Oskar Klein Centre for Cosmoparticle Physics, Dept.~of Physics, AlbaNova, Stockholm University, SE-10691 Stockholm, Sweden}

\date{\today}
\noaffiliation

\begin{abstract}
Neutrino-telescopes like Super-Kamiokande and IceCube have started to explore the neutrino fluxes from WIMP annihilations in the Sun. The non-observation of a signal can put constraints on the WIMP properties. We here focus on the neutrino signal from WIMP annihilation in the Sun and show that under reasonable assumptions, the non-observation of a signal from IceCube puts a much tighter constraint on the spin-dependent WIMP-proton scattering cross-section than current direct detection experiments like COUPP and KIMS. For the spin-independent scattering cross-section, the limits from IceCube and current direct detection experiments like XENON10 and CDMS place similar constraints. We here go through the assumptions being made and the uncertainties that arise in converting from limits on the muon flux from the Sun to limits on the WIMP-proton cross-section, and present our results as easy to use conversion factors.
\end{abstract}

\maketitle

\section{Introduction}

One of the favourite candidates for the dark matter in the Universe is Weakly Interacting Massive Particles (WIMPs), of which neutralinos that arise in supersymmetric extensions of the standard model is a popular example (see \textit{e.g.}\ Ref.~\cite{susy_dm} for a review). If WIMPs are the dark matter in the Universe, they will be abundant in the Milky Way halo where they can produce annihilation radiation (gamma rays, neutrinos, positrons, antiprotons, antideutrons, \ldots) that can be searched for with \textit{e.g.}\ the Pamela \cite{pamela} and Fermi \cite{fermi} satellites. The WIMPs can also scatter in heavy objects like the Sun \cite{sun-capture}, where they can lose enough energy to be captured gravitationally and eventually sink to the core. Once in the core, they can annihilate and produce \textit{e.g.}\ neutrinos that can be searched for with neutrino telescopes like Super-Kamiokande \cite{super-k}, IceCube \cite{icecube} and Antares \cite{antares}. Another route to search for the WIMPs is direct detection experiments where one tries to observe WIMP scatterings on a detector target. For both the neutrino signal from the Sun and the direct detection experiments, the strength of the signal depends crucially on either the spin-independent and/or the spin-dependent WIMP-nucleon scattering cross-section. Currently, XENON10 \cite{xenon10} and CDMS \cite{cdms} place the most stringent direct limits on the spin-independent scattering cross-section, whereas KIMS \cite{kims} and COUPP \cite{coupp} put the most stringent direct limits on the spin-dependent scattering cross-section. However, as the neutrino signal from the Sun as searched for by neutrino telescopes also depends strongly on these cross-sections, we can use these searches to put limits on the same scattering cross-sections. This is of course more involved than for direct detection experiments as we need to consider the capture in the Sun, the subsequent annihilation, production of neutrinos, propagation, interaction and oscillation of neutrinos, and finally scattering near the neutrino telescope to produce observable muons ($\mu^+$ and $\mu^-$). Most of these steps can be done in a fairly straightforward approach under reasonable assumptions, and the limit derived on the spin-dependent scattering cross-section is highly competitive with those from direct detection experiments. A similar analysis was done in Ref.~\cite{kjgs}, 
but those results need updating, mainly because neutrino oscillations change the neutrino fluxes, but also because better modelling of hadronization/decay of quark jets affect the results. We will here perform a more detailed study, including full three flavour neutrino oscillations. For this, we will use the WimpSim \cite{wimpsim} WIMP annihilation Monte Carlo as built into DarkSUSY~\cite{dsusy} 5.0.3, and derive conversion factors to go from limits on the muon flux from neutrino telescopes to limits on the spin-independent and spin-dependent WIMP-proton scattering cross-sections. At the end, we will apply these on the recent results from Super-Kamiokande \cite{super-k} and IceCube \cite{ic22-limits}.
The correlation between signals in direct detection experiments and neutrino telescopes was also investigated in Ref.\ \cite{ukv-dama}, in an analysis based on the results in \cite{kjgs}, but adding annual modulation to the picture.
In another recent paper \cite{hooper-low}, the correlation between neutrino telescopes and direct detection experiments was revisited for light WIMP dark matter (below 20 GeV). We will here be more general than in that paper and focus on a wide range of WIMP masses.

\section{WIMP capture and neutrinos}

As there are many assumptions needed to calculate the neutrino flux from the scattering cross-section, we will do this in two ways. First, we will focus on a `standard' calculation, with reasonable assumptions usually made in the literature. We will then go through some of these assumptions and perform a conservative calculation where we instead change our assumptions to see how low fluxes we can reasonably get for given scattering cross-sections.

\subsection{Standard calculation}

The number of neutralinos in the Sun, $N$, is described by the differential equation
\begin{equation}
 \frac{dN}{dt} = C_{C}-C_{A}N^{2}-C_{E}N,
\end{equation}
where the three constants describe capture ($C_{C}$), annihilation ($C_{A}$), and evaporation ($C_{E}$). For the masses of interest here, the evaporation is small and can be neglected \cite{evap}. Note that $C_{C}$ depends on the scattering cross-sections on the elements in the Sun, whereas $C_{A}$ depends on the annihilation cross-section.
The neutralino annihilation rate $\Gamma_{A}$ is then
\begin{align}
 \Gamma_{A}&\equiv\frac{1}{2}C_{A}N^{2} = \frac{1}{2}C_{C}\mathrm{tanh}^{2}(t/\tau) \label{eq:gammaa} \\
 \tau&\equiv1/\sqrt{C_{C}C_{A}}, \label{eq:tau}
\end{align}
where the present rate is found for $t=t^{\odot}\simeq4.5\cdot10^{9}$ years.
When $t^{\odot}/\tau \gg1$ annihilation and capture are in equilibrium, $\frac{dN}{dt}=0$, and 
\begin{equation}
 \Gamma_{A}=\frac{1}{2}C_{C}.
\end{equation}
Thus, in equilibrium, $\Gamma_{A}$ only depends on the capture rate $C_{C}$, \textit{i.e.}\ $\Gamma_{A}$ only depends on the scattering cross-sections, and not on the annihilation cross-section. From $\Gamma_{A}$ it is then straightforward (but tedious) to calculate the resulting muon flux at a neutrino telescope for a given WIMP annihilation channel.

For the capture ($C_{C}$) and annihilation ($C_{A}$) we follow the treatment in Ref.~\cite{gould} as implemented in DarkSUSY \cite{dsusy}. In short, we have to integrate the WIMP scatterings over the volume of the Sun and the velocity of the WIMPs in the galactic halo and will here review the most essential steps in this calculation.

For the velocity of the WIMPs, we will here assume that it follows a Maxwellian distribution, 
\begin{equation}
 \frac{f(u)}{u}=\sqrt{\frac{3}{2\pi}}\frac{n_{\chi}}{v_{d}\cdot v_{\odot}}\left(\exp\left(-\frac{3(u-v_{\odot})^{2}}{2v_{d}^{2}}\right) - \exp\left(-\frac{3(u+v_{\odot})^{2}}{2v_{d}^{2}}\right) \right),
\end{equation}
where $u$ is the velocity of the WIMP (outside the potential well of the Sun), $v_{\odot}$ = 220 km/s is the velocity of the Sun relative to the halo, $v_{d}$ = 270 km/s is the WIMP velocity dispersion, and $n_{\chi}$ is the WIMP number density. We assume that the local WIMP density is 0.3 $\mathrm{GeV/cm^{3}}$. At a given interaction point in the Sun, the WIMP velocity is given by $w=\sqrt{u^{2}+v^{2}}$, where $v$ is the escape velocity at that point.

The WIMP capture rate per unit shell volume from element $i$ in the Sun is \cite{gould}
\begin{equation}
 \frac{dC_{i}}{dV}=\int_{0}^{u_{max}}\mathrm{d}u\frac{f(u)}{u}w\Omega_{v,i}(w),
\label{eq:dCdV}
\end{equation}
where $\Omega_{v,i}(w)$ is the capture probability per unit time for element $i$. In Eq.~(\ref{eq:dCdV}) we integrate up to the velocity $u_{max}=2v\sqrt{\mu}/(\mu-1)$ at which the WIMPs scatter to the escape velocity $v$. Here $\mu \equiv m_{\chi}/m_{i}$, for the WIMP mass $m_{\chi}$ and the mass $m_{i}$ of element $i$.

For scattering on heavier elements than Hydrogen, we also need to take decoherence into account by introducing a form factor suppression. Following standard lore, we assume a Helm-Gould exponential nuclear form factor \cite{gould} for the momentum transfer $q$ on element $i$
\begin{equation}
 |F_{i}(q^{2})|^{2} = \exp(-\Delta E/E_{i}^{0}) \quad ; \quad \Delta E = \frac{q^2}{2m_\chi}
\end{equation}
where
\begin{align}
\label{eq:HG}
 E_{i}^{0}&=\frac{{×}3\hbar^{2}}{2m_{\chi}R_{i}^2},\\ \nonumber
 R_{i}&=\left[0.91(m_{i}/\mathrm{GeV})^{1/3}+0.3\right] \cdot 10^{-15} \mbox{~m}.
\end{align}
$\Omega_{v,i}(w)$ can then be calculated analytically and we arrive at the expression \cite{gould}
\begin{equation}
 \Omega_{v,i}(w) = \sigma_{\chi i}\cdot n_{i}\frac{(\mu+1)^{2}}{2\mu}E_{i}^{0}\left[\mathrm{exp}\left(-\frac{m_{\chi}u^{2}}{2E_{i}^{0}}\right)-\mathrm{exp}\left(-\frac{2\mu}{(\mu+1)^{2}}m_{\chi}\frac{u^{2}+v^{2}}{E_{i}^{0}}\right)\right].
\end{equation}
Here $\sigma_{\chi i}$ is the WIMP scattering cross-section on element $i$, and $n_{i}$ is the number density of element $i$ in the Sun.

To get the total capture rate we integrate over the radius of the Sun, $R_{\odot}$, and finally arrive at \cite{gould}
\begin{equation}
 C_{C}=\int_{0}^{R_{\odot}}4\pi r^{2}\mathrm{d}r \sum_{i}\frac{\mathrm{d}C_{i}(r)}{\mathrm{d}V},
\label{eq:c_tot}
\end{equation}
where the solar composition is taken from Ref.~\cite{solar_model}, where we take the solar model BS2005-OP as our default choice. For elements heavier than Oxygen, the solar model BS2005-OP only gives the total abundance and we use the relative abundances as given in Ref.~\cite{gresau98} for these. In total, we include the 16 most important elements up to Nickel.

We can calculate the spin-independent (SI) cross section on a nucleus with mass number $A$, $\sigma_{\chi A}^{SI}$, as \cite{susy_dm} 
\begin{equation}
 \sigma_{\chi A}^{SI}=\sigma^{SI}A^{2}\frac{(m_{\chi}m_{A})^{2}}{(m_{\chi}+m_{A})^{2}}\frac{(m_{\chi}+m_{p})^{2}}{(m_{\chi}m_{p})^{2}}
\label{eq:chia}
\end{equation}
where $\sigma^{SI}$ is the SI WIMP-proton scattering cross-section.

For the spin-dependent (SD) WIMP-proton cross-section $\sigma^{SD}$, Hydrogen is the most important element in the Sun, and we can neglect the heavier elements in this case. 

The muon flux ($\mu^{-}$ and $\mu^{+}$) at the detector is calculated from the WIMP annihilation rate in the Sun as
\begin{equation}
 \Phi_{\mu}=\frac{\Gamma_{A}\cdot n}{4\pi D_{\odot}^{2}}\int_{E_{\mu}^{th}}^{\infty}\mathrm{d}E_{\mu}\int_{E_{\mu}^{th}}^{\infty}\mathrm{d}E_{\nu} \int_{0}^{\infty}\mathrm{d}\lambda \int_{E_{\mu}}^{E_{\nu}} \mathrm{d}E_{\mu}^{'}P(E_{\mu},E_{\mu}^{'},\lambda)\frac{\mathrm{d}\sigma_{\nu}(E_{\nu},E_{\mu}^{'})}{\mathrm{d}E_{\mu}^{'}}\sum_{i}P(\mu,i)\sum_{f}B_{f}\frac{\mathrm{d}N_{i}^{f}}{\mathrm{d}E_{\nu}},
\label{eq:muflux}
\end{equation}
where $D_{\odot}$ is the distance from the center of the Sun, $n$ the target number density, $E_{\mu}^{th}$ the detector's energy threshold, $\lambda$ the muon range, $P(E_{\mu},E_{\mu}^{'},\lambda)$ the probability for a muon of energy $E_{\mu}^{'}$ to have a final energy $E_{\mu}$ after a path-length $\lambda$ in the detector material, $\mathrm{d}\sigma_{\nu}(E_{\nu},E_{\mu}^{'})/\mathrm{d}E_{\mu}^{'}$ is the differential neutrino cross-section for production of a muon with energy $E_{\mu}^{'}$ from a neutrino of energy $E_{\nu}$, $P(\mu,i)$ is the probability that a produced neutrino of flavour $i$ oscillates to flavour $\mu$ in the detector, $B_{f}$ is the branching ratio for annihilation channel $f$, and $\mathrm{d}N_{i}^{f}/\mathrm{d}E_{\nu}$ is the differential number of neutrinos of flavour $i$ produced per annihilation in channel $f$. We perform this calculation with WimpSim \cite{wimpsim} including interactions, neutrino oscillations and regeneration of neutrinos from tau decay. We have performed these simulations with the neutrino oscillation parameters \cite{Maltoni} (see \cite{wimpsim} for details)
%
$\theta_{12} = 33.2^\circ$,
$\theta_{13} = 0^\circ$,
$\theta_{23} = 45.0^\circ$,
$\delta = 0$,
$\Delta m_{21}^2 = 8.1\times 10^{-5} \, {\rm eV}^2$ and
$|\Delta m_{31}^2| = 2.2\times 10^{-3} \, {\rm eV}^2$
but our results are not very sensitive to the exact choice of these parameters. The results of these WimpSim simulations are included in DarkSUSY \cite{dsusy} as easy to use tables and interpolation routines. For our purpose here, we use a muon energy threshold of 1 GeV as this is how most neutrino telescopes report their results. We will also use the muon fluxes without any angular cut on the muons.

\begin{figure}[th]
\includegraphics[width=0.65\linewidth]{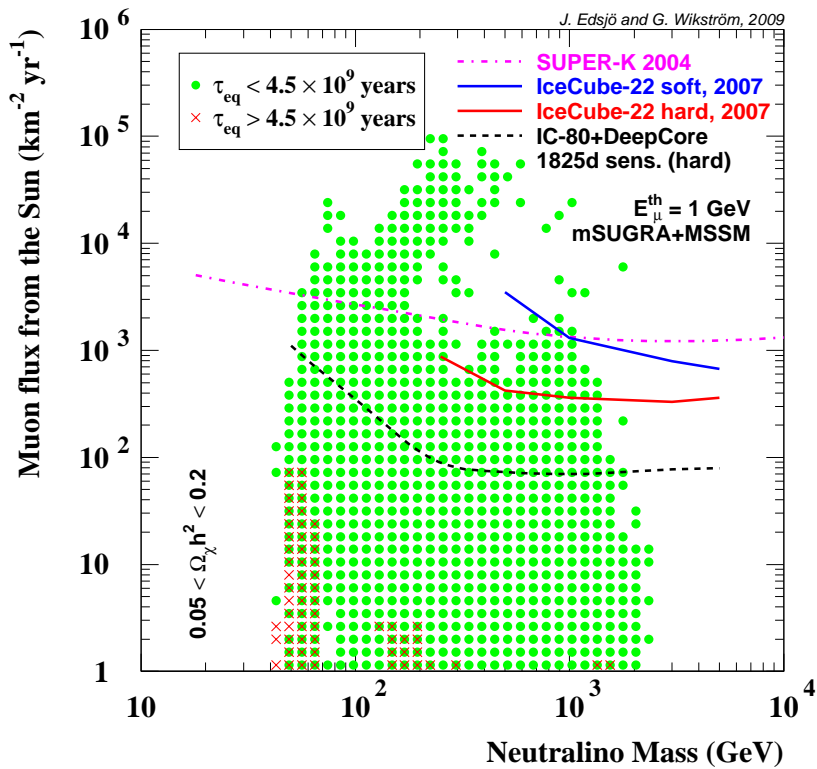}
\caption{Muon flux from the Sun as a function of neutralino mass, for mSUGRA and MSSM models where neutralinos are (dots) or are not (crosses) in equilibrium in the Sun. Also shown are limits from IceCube \cite{ic22-limits}, and Super-Kamiokande \cite{super-k}, and the full IceCube estimated sensitivity \cite{ic22-limits}. The muon flux is given above a threshold of 1 GeV.}
\label{fig:equilibrium}
\end{figure}

We then have all the necessary ingredients to calculate the flux of muons (or neutrinos) in a neutrino telescope from any given WIMP scattering and annihilation cross-sections. However, for the neutrino telescope limits to be really competitive, we need to have a situation where we have equilibrium between annihilation and capture as discussed above (\textit{i.e.}\ we need to have $\tau_{\rm eq} < 4.5\cdot 10^9$ years). If we do not have equilibrium, neutrino telescope limits will still constrain WIMP models, but we lose the direct correlation between the neutrino-induced muon fluxes and the scattering cross-sections. To investigate this further, we need to look at specific WIMP models and we will here focus on the neutralino that arises as a natural WIMP dark matter candidate in supersymmetric extensions of the standard model for particle physics, like the Minimal Supersymmetric Standard Model (MSSM). We have made extensive scans of the MSSM model parameters both in mSUGRA models and low-energy phenomenological MSSM-7 and MSSM-9 models (in these models, we specify the supersymmetric parameters at the electroweak scale instead of at the GUT scale as in mSUGRA, see \cite{ibeplus} for details). In Fig.~\ref{fig:equilibrium} we show the neutrino-induced muon flux $\Phi_\mu$ for models passing current accelerator constraints and with a relic density in the cosmologically preferred range (generously chosen), together with recent limits from IceCube \cite{ic22-limits}, and Super-Kamiokande \cite{super-k}, and the full IceCube estimated sensitivity \cite{ic22-limits}. In this figure we indicate for which models we do have equilibrium between capture and annihilation and for which we do not. For all models with high fluxes that are currently probed and will be probed in the near future by neutrino telescopes, equilibrium is fulfilled and we will from now on assume that we do have equilibrium. For hypothetical scenarios where this assumption would be invalid, it is straightforward to rescale our results with the $\tanh^2(t/\tau_{\rm eq})$ factor given in Eq.~(\ref{eq:gammaa}) above. We also note that for another popular class of models, Kaluza-Klein dark matter in Universal Extra Dimensions (UED) \cite{ued}, we also always have equilibrium or near equilibrium between capture and annihilation. Na\"{\i}vely, we do expect to have equilibrium for models that can be probed with currently planned experiments, as the annihilation cross-section that enters $\tau$ in Eq.~(\ref{eq:tau}) is constrained by the relic density constraint to be of the order of $\langle \sigma v \rangle_{\rm ann} \simeq 3 \cdot 10^{-27}$ cm$^3$ s$^{-1}$. The above examples for MSSM and UED confirm that this is the case for these specific models.

\subsection{Conservative calculation and uncertainties}

The calculation above includes `standard' assumptions both from particle physics and astrophysics. However, we want to use the neutrino telescope results to put a limit on the scattering cross-section. Hence, we need to investigate processes that can affect our results. In particular we want to investigate the processes that can reduce the neutrino fluxes since a reduction would mean that our limits on the scattering cross-sections would be increase (\textit{i.e.}\ they would have been too optimistic). 

First of all, there are some uncertainties in the solar model, especially regarding the abundances of heavy elements. In a conservative setup, we instead focus on the solar model BS2005-AGS,OP \cite{solar_model} which includes new lower estimates of the abundances of the heavier elements (for the actual relative abundance we use the estimates in \cite{gresau98} for heavier elements than Oxygen). This will not affect capture via spin-dependent scattering as this occurs on Hydrogen, whose abundance is well-known. It will however reduce capture via spin-independent scattering, especially for heavier WIMPs (heavier than a few hundred GeV), where capture on the heavier elements is important.

Another effect that can reduce the capture rate is gravitational effects from the planets, especially Jupiter, during the capture process \cite{jupiter1,jupiter2}. In Ref.\ \cite{jupiter1,jupiter2}, it was estimated that for low scattering cross-sections, the life time of WIMP orbits after the first scattering in the Sun is quite long (\textit{i.e.}\ it takes a long time before they scatter again and eventually sink to the core). If the WIMP orbit reaches out to Jupiter, there is a large probability that Jupiter will affect the orbit and disturb it so that it no longer passes through the Sun and eventually throws the WIMP out of the solar system back into the Milky Way halo. To get a conservative estimate of this effect, we have modified the capture expressions above so that WIMPs that reach out to Jupiter after the first scatter in the Sun will be considered as not captured. This reduces capture, especially for heavy WIMPs and especially for scattering on light elements in the Sun (\textit{i.e.}\ it is more pronounced for spin-dependent scattering). For 1 (10) TeV WIMPs the reduction is about 13\% (87\%) for spin-dependent capture and about 1\% (12\%) for spin-independent capture.

The form factors introduce another source of uncertainty in the spin-independent capture, and the usually assumed Helm-Gould form factors are rather approximate (but convenient to use). In Ref.\ \cite{form-factors}, the effect of different form factors on direct detection rates was investigated and the conclusion was that the total scattering rate could change by up to about 20\% with more accurate form factors. For our conservative estimate, we will here use the Helm-Lewin-Smith form factors \cite{lewin-smith,form-factors}, which for mass number $A$ are given by

\begin{align}
 R&=\sqrt{ \left( c^{2}+(7/3)\pi^{2}a^{2}-5s^{2} \right)} \\ \nonumber
s&=0.9~\cdot 10^{-15} \mbox{~m},~a\simeq0.52~\cdot 10^{-15} \mbox{~m},~c\simeq[1.23A^{1/3}-0.60]~\cdot 10^{-15} \mbox{~m}.
\end{align}

Another source of uncertainty is of course the local dark matter density and velocity distribution. For our conservative estimate, we keep the local halo density fixed to $\rho_0 = 0.3$ GeV/cm$^3$ as for the standard calculation, as it is easy to rescale our results for other densities. The local WIMP density also enters in the same way for direct detection experiments, so it does not affect the conversion factors below. A bigger concern is the velocity distribution as direct detection experiments are more sensitive to high velocity WIMPs (as they produce larger energy nuclear recoils), whereas the capture process in the Sun is more efficient for low velocity WIMPs. N-body simulations indicate (see \textit{e.g.}\ \cite{n-body}) that dark matter halos probably are both anisotropic and deviates from the simple Maxwellian form. However, dynamical constraints \cite{vel-dyn} indicate that they cannot change by too much and still be consistent with observations. It seems reasonable to assume though, that the velocity distribution is not that different from a Maxwellian distribution, that it would reduce the capture in the Sun by more than a factor of two. For our conservative estimate, we have though not included the effects of changing the velocity distribution, as there is yet no consensus on how different it could be and still be compatible with observations.  There is also a recent proposal \cite{read} that there might be a dark matter disk in the Milky Way in addition to the regular dark matter halo. This disk could have a local density of (0.25--1.5) $\rho_0$ and would corotate with the stars in the galaxy. The velocity distribution of this population would therefore be peaked at much lower velocities than the regular halo, which would cause very little change in direct detection experiments (except if the energy thresholds could be reduced), but would increase the capture rate in the Sun by maybe one order of magnitude \cite{bruch}. The existence of this dark matter disk is still debated and as it increases the capture rate instead of decreasing it, we do not include it in our conservative estimate. We just want to point out that if such a disk exists, the neutrino telescopes could be even more sensitive than the standard calculation gives.

\section{Results}

We are now ready to calculate our results and will first focus on the conversion rates from muon flux limits to scattering cross-section limits. We will then compare with earlier calculations and at the end we will apply our conversion factors on the current best limits from neutrino telescopes and compare with direct detection experiments.

\subsection{Conversion factors}

\begin{figure}[th]
 \includegraphics[width=0.49\linewidth]{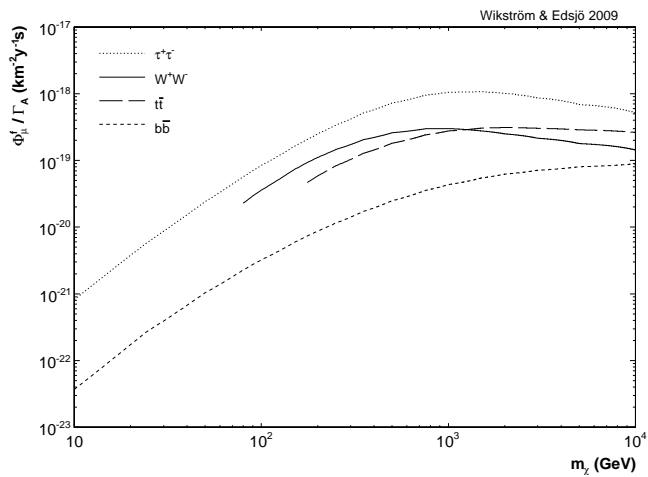}
\caption{The conversion factor
  $\frac{\Phi_{\mu}^{f}/\mathrm{km}^{-2}y^{-1}}{\Gamma_{A}/\mathrm{s}^{-1}}$
  as a function of $m_{\chi}$, for annihilation channels $f=W^{+}W^{-},~\tau^{+}\tau^{-},~t\overline{t},~\mathrm{and}~b\overline{b}$. This conversion is identical in the standard and conservative calculations.}
\label{fig:fovera}
\end{figure}

If we assume that one annihilation channel dominates (\textit{i.e.}\ assuming $B_{f}=1$ for some annihilation channel $f$ in Eq.~(\ref{eq:muflux})) we can relate the observed muon flux $\Phi_{\mu}^{\textit{obs}}$ to the annihilation rate
\begin{equation}
 \Phi_{\mu}^{\textit{obs}}=\Phi_{\mu}^{f}=\eta^{f}(m_{\chi})\Gamma_{A},
\end{equation}
using the functions $\eta^{f}(m_{\chi})$, shown for four (extreme) channels in Fig.~\ref{fig:fovera}. These values are the same in the standard and conservative calculations. Note that for a specific WIMP model, we typically know the branching fractions $B_f$, but for general WIMPs, we do not have this knowledge and in extreme cases the uncertainty coming from this lack of knowledge could be about an order of magnitude as seen in the Fig.~\ref{fig:fovera}. However, for most WIMP models, the uncertainty is usually much smaller than this. For supersymmetric neutralinos, the branching fraction to hard channels is typically $>10\%$ for $m_{\chi}>m_{W}$, so the sum of $\eta$'s is typically rather close (within a factor of a few) to the values for a hard channel like $W^+W^-$. For Kaluza-Klein dark matter in models of Universal Extra Dimensions, the branching fractions are rather well-known and with a large hard contribution. Hence, for most realistic models, $\eta$ is closer to the hard channels like $W^+W^-$ than the soft ones like $b\bar{b}$. To get a really conservative limit, we could use the soft $b\bar{b}$ channel as an extreme case, but we have here chosen to present results for these different channels separately.

\begin{figure}[th]
\includegraphics[width=0.49\linewidth]{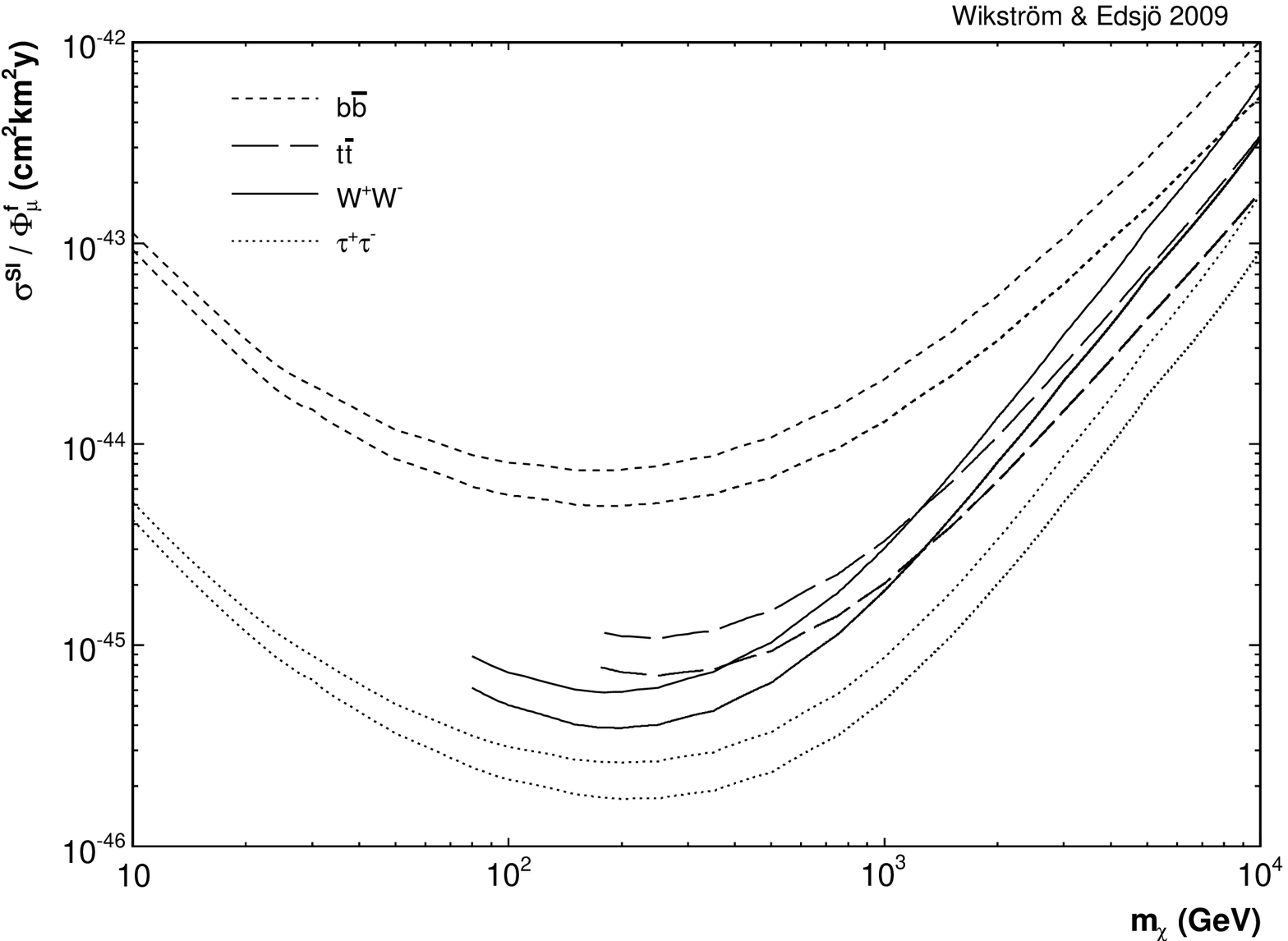}
\hfill
\includegraphics[width=0.49\linewidth]{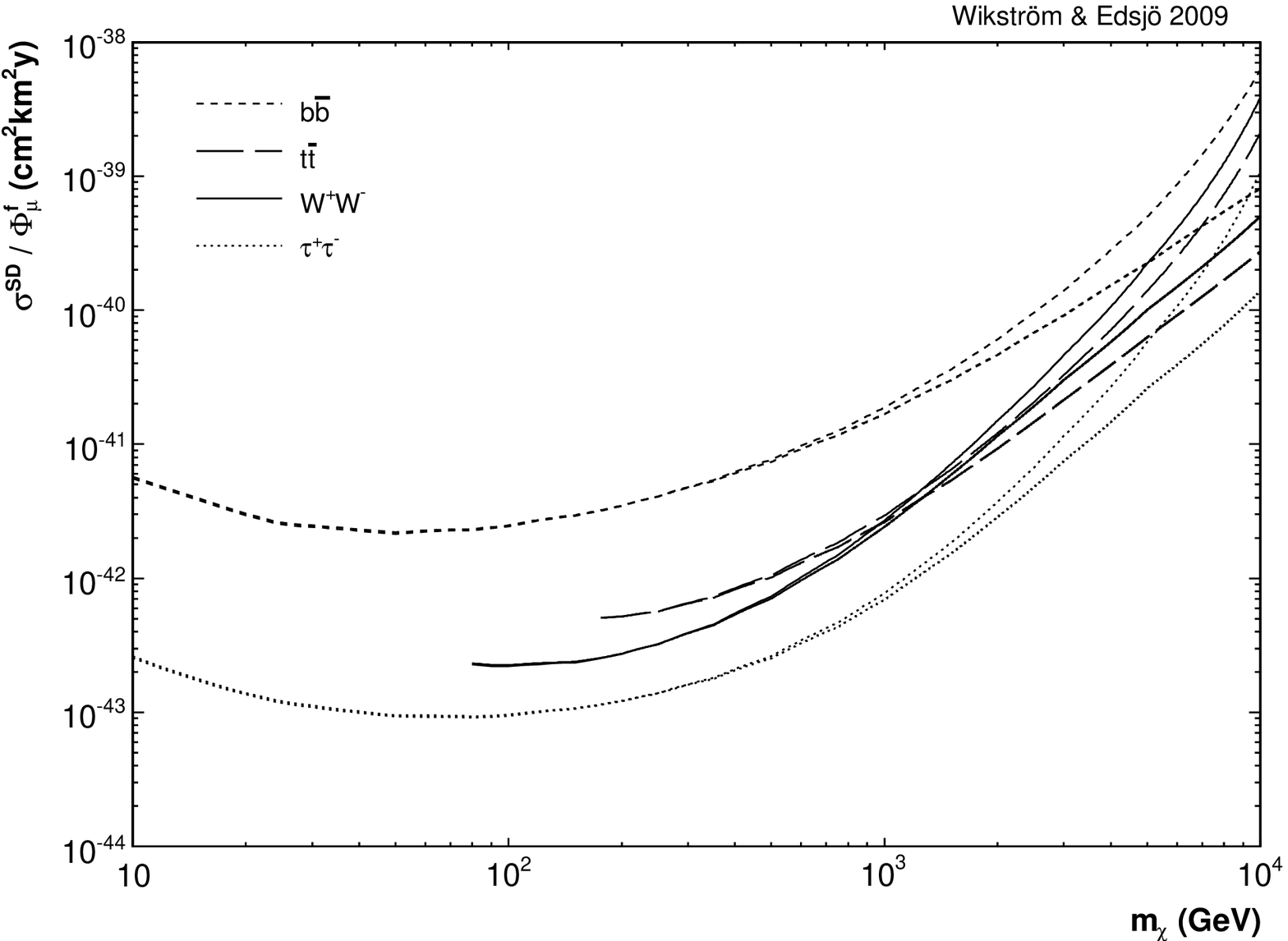}
\caption{The conversion factors
  $\kappa_f^{SI}=\frac{\sigma^{SI}/\mathrm{cm}^{2}}{\Phi_{\mu}^{f}/\mathrm{km}^{-2}y^{-1}}$
  (left panel) and
  $\kappa_f^{SD}=\frac{\sigma^{SD}/\mathrm{cm}^{2}}{\Phi_{\mu}^{f}/\mathrm{km}^{-2}y^{-1}}$
  (right panel) as a function of $m_{\chi}$, for annihilation channels $f=W^{+}W^{-},~\tau^{+}\tau^{-},~t\overline{t},~\mathrm{and}~b\overline{b}$. Lower lines for each channel are from the standard calculation and upper lines are from the conservative calculation. For the muon flux we have used a threshold of 1 GeV.}
\label{fig:soverf}
\end{figure}

Further on, to get a conservative limit on either the spin-independent or spin-dependent scattering cross-section, we assume that either one of these dominates capture. We can then through Eqs.~(\ref{eq:c_tot}) and (\ref{eq:chia}) get the WIMP-proton scattering cross-sections $\sigma^{SI}$ or $\sigma^{SD}$ as a function of $\Gamma_{A}$
\begin{align}
 \sigma^{SI}&=\lambda^{SI}(m_{\chi})\Gamma_{A}\\ \nonumber
 \sigma^{SD}&= \lambda^{SD}(m_{\chi})\Gamma_{A},
\end{align}
for functions $\lambda^{SI}(m_{\chi})$ and $\lambda^{SD}(m_{\chi})$.
We can then finally relate the cross-section to the muon flux, as 
\begin{align}
\sigma^{SI}&=\frac{\lambda^{SI}(m_{\chi})}{\eta^{f}(m_{\chi})}\Phi_{\mu}^{f}\equiv\kappa_{f}^{SI}(m_{\chi})\Phi_{\mu}^{f} \\ \nonumber
\sigma^{SD}&=\frac{\lambda^{SD}(m_{\chi})}{\eta^{f}(m_{\chi})}\Phi_{\mu}^{f}\equiv\kappa_{f}^{SD}(m_{\chi})\Phi_{\mu}^{f}.
\end{align}
The conversion factors $\kappa_{f}^{SI}(m_{\chi})$ and $\kappa_{f}^{SD}(m_{\chi})$ between cross-section and muon flux are shown for four annihilation channels in Fig.~\ref{fig:soverf} for the standard and conservative calculations. For the SI case, we note that the main effect of the conservative calculation is to increase $\kappa^{SI}$ due to the different form factor. At higher masses, the lower abundances of heavy elements and the effect of Jupiter increases slightly further. For the SD case, we see essentially no differences between the two calculations at low masses as the only significant difference between the calculations in this case is the reduction of capture due to Jupiter, that starts increasing $\kappa^{SD}$ at high masses.

The conversion factors in Fig.~\ref{fig:soverf} can be used to convert a muon flux limit from a neutrino telescope to a limit on the SI or SD WIMP-nucleon scattering cross-section. We have also developed a web tool \cite{webtool} that can be used to get accurate values for these conversion factors for any WIMP mass between 10 GeV and 10 TeV\@. The web tool also contains more annihilation channels than those in the figure and also allows to convert from fluxes with different muon energy thresholds and angular cuts than those assumed here. The web tool is also available as a downloadable Fortran program to be used with DarkSUSY \cite{dsusy}.

\subsection{Comparison with earlier calculations}

\begin{figure}[th]
\centerline{\includegraphics[width=0.65\linewidth]{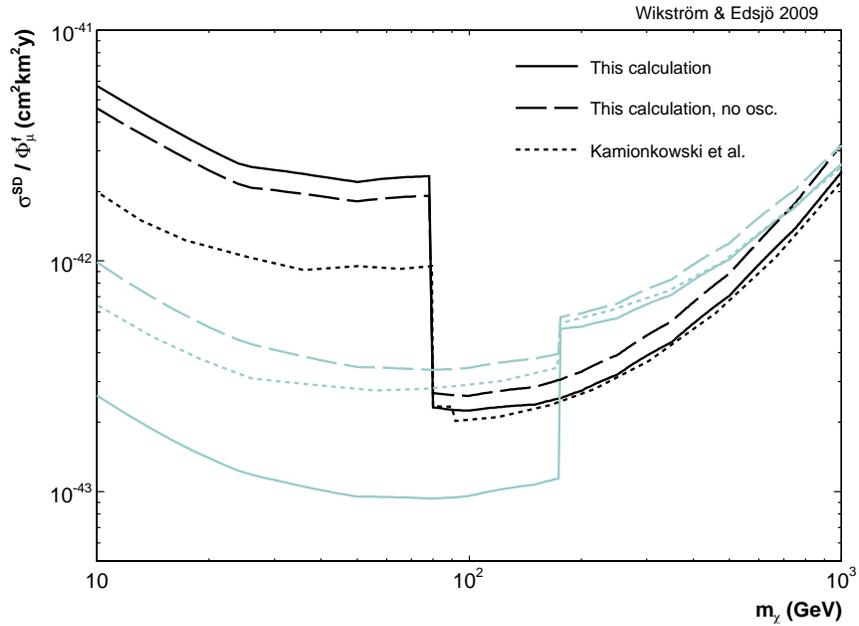}}
\caption{The conversion factors $\kappa_f^{SD}=\frac{\sigma^{SD}/\mathrm{cm}^{2}}{\Phi_{\mu}^{f}/\mathrm{km}^{-2}y^{-1}}$ calculated from values given by Kamionkowski \textit{et al.} \cite{kjgs} (dotted), compared with this calculation including (solid) and neglecting (dashed) neutrino oscillations. See text for details.}
\label{fig:kjgs}
\end{figure}

We can now compare our results with earlier estimates. In Kamionkowski \textit{et al.} \cite{kjgs}, the event rates in neutrino telescopes were compared with the expected event rates in a hypothetical Hydrogen direct detection detector (and were given in 
units of $\mathrm{m^{2}kg^{-1}}$). They assumed a set of different branching fractions $B_f$ for different mass ranges and gave their result as the spin-dependent scattering rate. Using the same expression for the direct rate, given in Ref.~\cite{directrate}, we have calculated the equivalent conversion factors $\kappa_{f}^{SD}(m_{\chi})$ in terms of the cross-section instead of the direct detection event rate. In Fig.~\ref{fig:kjgs}, we show these conversion factors from Ref.~\cite{kjgs} (dotted line) and compare with our results with our standard calculation (solid line). As the calculation in Ref.~\cite{kjgs} does not include neutrino oscillations, for comparison reasons we also show our results without including neutrino oscillations (dashed line). 

In this plot, the branching fractions for the black lines are assumed to be $f=b\overline{b}$ for $m_{\chi} < m_{W}\simeq 80~\mathrm{GeV}$ and $f=W^{+}W^{-}$ or $Z^{0}Z^{0}$ for $m_{\chi} > 80~\mathrm{GeV}$. The grey lines are instead for $f=\tau^{+}\tau^{-}$ for $m_{\chi} < m_t \simeq 175~\mathrm{GeV}$ and $f=t\overline{t}$ for $m_{\chi} > m_t \simeq 175~\mathrm{GeV}$. In this figure, we see some discrepancies with the calculation by Kamionkowski \textit{et al.}. 

First of all, our conversion factors for $b\bar{b}$ are higher than their results, which is mostly due to a different treatment of hadronization and decay of the $b$ quarks. We have used Pythia \cite{pythia} for hadronization/decay, whereas \cite{kjgs} have used analytical expressions that \textit{e.g.}\ do not include final state radiation. Also, the treatment of $b$ meson interactions in the Sun is slightly different in our treatment where we have used updated interaction cross-sections (see \cite{wimpsim} for details). Our results for $b\bar{b}$ agree well with those in Ref.~\cite{cirelli}, and as they are softer than in \cite{kjgs} our conversion factors are higher. For $W^+W^-$ and $t\bar{t}$, our results agree fairly well, but for $\tau^+\tau^-$, we see a large discrepancy between the grey solid and dashed curves (below $m_t$). This discrepancy comes from neutrino oscillations not being included in the \cite{kjgs} treatment and our results with neutrino oscillations turned off (grey dashed curve), agree fairly well with the results in \cite{kjgs}. The reason neutrino oscillations have such a big effect on the $\tau^+\tau^-$ channel is that we get significantly more tau neutrinos than other neutrino flavours from the $\tau$ decays and oscillations will efficiently transform these to detectable muon neutrinos at the detector. Hence, the neutrino flux is significantly enhanced due to neutrino oscillations, which will result in a lower conversion factor in Fig.~\ref{fig:kjgs}.

Other smaller differences come from us using an updated solar model and the use of the full expressions for solar capture (including an integration over the solar radius).

\begin{figure}[th]
\centerline{\includegraphics[width=0.65\linewidth]{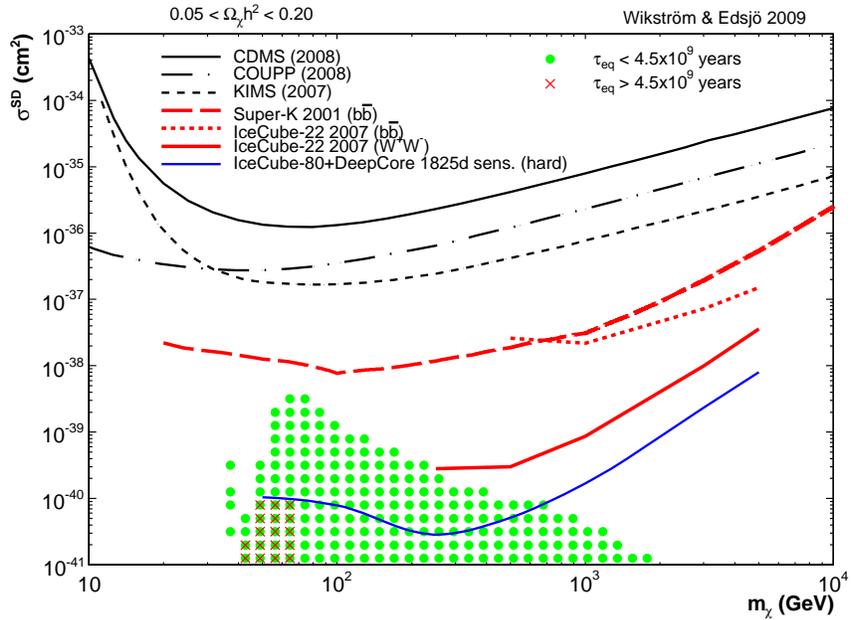}}
\caption{Limits on spin-dependent WIMP-proton cross-section as a function of neutralino mass, for Super-K calculated from the muon flux limit \cite{super-k}, and for IceCube-22 \cite{ic22-limits}. Limits from CDMS \cite{cdms}, COUPP \cite{coupp}, KIMS \cite{kims}, and the full IceCube estimated sensitivity \cite{ic22-limits} are shown for comparison. Also shown are mSUGRA and MSSM models where neutralinos are (dots) or are not (crosses) in equilibrium in the Sun}
\label{fig:limits}
\end{figure}

We have also compared with the calculation in Ref.~\cite{hooper-low}, where they focus on WIMP masses up to 20 GeV\@. They have re-evaluated the Super-Kamiokande limits for this low-mass case and used an updated method of the one in \cite{kjgs}. Their limits on the cross-section are lower than those we have here, where we use the Super-Kamiokande published results (see Fig.~\ref{fig:limits}). In \cite{hooper-low} they analyze the Super-Kamiokande data up to $30^\circ$ from the Sun, where there happens to be a downwards fluctuation in the number of recorded events compared to the background expectation, which results in a better limit. They have also used approximate expressions for WIMP capture in the Sun and similar analytic expressions for the neutrino yield from $b\bar{b}$ as in \cite{kjgs}. These effects taken together can explain the difference between our results.

\subsection{Limits on scattering cross-sections}

Using our conversion factors in Fig.~\ref{fig:soverf} (we here use the standard calculation), we can now convert limits on the muon flux from neutrino telescopes to limits on the scattering cross-section. As expected, the limits on the spin-independent scattering cross-section are are similar to, but not competitive with, current direct detection limits and we will instead focus on the spin-dependent scattering cross-sections. In Fig.~\ref{fig:limits} we show the limits on the spin-dependent scattering cross-section for our standard calculation using the recent muon flux limits from the IceCube 22-string detector \cite{ic22-limits} and the full IceCube estimated sensitivity \cite{ic22-limits}. For comparison, we also show the limits from Super-Kamiokande \cite{super-k}. For Super-Kamiokande, we convert their muon flux limit from a threshold of 1.6 GeV to limits on the flux with a 1 GeV threshold. They have also applied an angular cut in the range of $1.5^{\circ}$ to $19^{\circ}$ \cite{angular_sel} for different masses, and we also correct for this effect to get a limit on the full muon flux without angular cuts. The limits are further given for a mixture of $80\%~b\overline{b},~10\%~c\overline{c},~10\%~\tau^{+}\tau^{-}$, and we have here chosen to calculate the limit for $b\overline{b}$. In practice, the choice of annihilation channel mixtures in the Super-Kamiokande analysis mainly affects the choice of angular cuts, and as we have corrected for that, it is a good approximation to treat the limits as limits on the flux for any annihilation channel (the limits on the cross-section will depend on the annihilation channel though through the channel dependence in $\kappa^{SD}_{f}$). \footnote{For the Super-Kamiokande limits, there is a discrepancy between our calculated limits on $\sigma^{SD}$ and the limits Super-Kamiokande themselves have calculated \cite{super-k} on $\sigma^{SD}$ (by using the results in \cite{kjgs}). Their limits are about a factor of 5 higher and this difference cannot be explained with our (small) differences with the results in \cite{kjgs}. It probably stems from an inconsistent use of different direct detection event rate calculations when using the results in \cite{kjgs}. We have been in contact with the authors of \cite{super-k} about this possible error.}

For comparison, in Fig.~\ref{fig:limits}, we also show the direct detection limits on the spin-dependent scattering cross-section from CDMS \cite{cdms}, COUPP \cite{coupp}, and KIMS \cite{kims}. As can be seen, the neutrino telescope limits are highly competitive, even if we take the annihilation channel uncertainties into account. Even if we would switch to the conservative calculation, the limits from neutrino telescopes would be far better than current direct detection experiments.

\section{Conclusions}

We have calculated the conversion factors between muon fluxes in neutrino telescopes and direct detection scattering cross-sections and applied these to the current neutrino telescope limits from IceCube and Super-Kamiokande. The limits on the spin-dependent scattering cross section are shown to be highly competitive with current direct detection experiments.

The main difference between our result and the previous result of Kamionkowski \emph{et al.} \cite{kjgs} comes from including neutrino oscillation effects in the Sun. For $f=b\overline{b}$, there is another difference, mainly coming from how $b$ hadronization/decay and $b$-meson interactions are treated. In addition there is a smaller difference coming from a more exact calculation of the capture in the Sun in our treatment.

Finally, using these new conversion factors, experiments like IceCube, Super-Kamiokande and Antares can easily convert their limits on the muon flux to limits on the scattering cross sections (as was recently done by IceCube in \cite{ic22-limits}).

\section*{Acknowledgements}
We wish to thank CCAPP at Ohio State University, the IceCube WIMP search group, and Jan Conrad for valuable discussions. J.E. thanks the Swedish Research Council (VR) for support.



\begin{thebibliography}{99}
\bibitem{susy_dm} G.~Jungman, M.~Kamionkowski and K.~Griest, Phys. Rep. \textbf{267} (1996) 195.
\bibitem{pamela} P.~Picozza \textit{et al.}, Astropart. Phys. \textbf{27} (2007) 296.
\bibitem{fermi} A.~A.~Moiseev \textit{et al.}, Nucl. Instr. Meth. A \textbf{588} (2008) 41.
\bibitem{sun-capture} W.~H.~Press and D.~N.~Spergel, Astrophys. J \textbf{296} (1985) 679; J.~Silk, K.~Olive and M.~Srednicki, Phys.\ Rev.\ Lett.\ {\bfseries 55} (1985) 257.
\bibitem{super-k} S.~Desai \textit{et al.}, Phys. Rev. D \textbf{70} (2004) 083523.
\bibitem{icecube} A.~Achterberg \textit{et al.}, Astropart. Phys. \textbf{26} (2006) 155.
\bibitem{antares} J.~A.~Aguilar \textit{et al.} Astropart. Phys. \textbf{26} (2006) 314.
\bibitem{xenon10} J.~Angle \textit{et al.}, Phys. Rev. Lett. \textbf{100} (2008) 021303.
\bibitem{cdms} Z.~Ahmed \textit{et al.}, astro-ph/0802.3530.
\bibitem{kims} H.~S.~Lee \textit{et al.}, Phys. Rev. Lett. \textbf{99} (2007) 091301.
\bibitem{coupp} E.~Behnke \textit{et al.}, Science \textbf{319} (2008) 933.
\bibitem{kjgs} M.~Kamionkowski \textit{et al.}, Phys. Rev. Lett. \textbf{74} (1995) 5174.
\bibitem{wimpsim} M.~Blennow, J.~Edsj\"{o}, T. Ohlsson, JCAP \textbf{01} (2008) 021.
\bibitem{dsusy} P.~Gondolo \textit{et al.}, JCAP \textbf{0407} (2004) 008.
\bibitem{ic22-limits} R.~Abbasi \textit{et al.}, astro-ph/0902.2460.
\bibitem{ukv-dama} P.~Ullio, M.~Kamionkowski and P.~Vogl, JHEP \textbf{0107} (2001) 044. [arXiv: hep-ph/0010036]

\bibitem{hooper-low}  D.~Hooper \textit{et al.}, astro-ph/0808.2464.
\bibitem{evap} K.~Griest and D.~Seckel, Nucl. Phys. B \textbf{283} (1987) 681.
\bibitem{gould} A.~Gould, Astrophys. J. \textbf{321} (1987) 571.
\bibitem{solar_model} J.~Bahcall \textit{et al.}, astro-ph/0412440.
\bibitem{gresau98} N.~Grevesse and A.J.~Sauval, Space Science Review {\bf 85} (1998) 161.
\bibitem{Maltoni} M.~Maltoni, T.~Schwetz, M.A.~T{\'o}rtola, and J.W.F.~Valle, New J.\ Phys.\ {\bfseries 6} (2004) 122.
\bibitem{ibeplus} L.~Bergstr\"{o}m, T.~Bringmann and J.~Edsj\"{o}, Phys.\ Rev.\ {\bf D78} (2008) 103520. [arXiv: 0808.3725]
\bibitem{ued} D.~Hooper and G.~D.~Kribs, Phys.\ Rev.\ {\bf D67} (2003) 055003. [arXiv: hep-ph/0208261]
\bibitem{jupiter1} A.~Peter and S.~Tremaine, [arXiv:0806.2133].
\bibitem{jupiter2} A.~Peter, [arXiv:0902.1347].
\bibitem{form-factors} G.~D\={u}da, A.~Kemper, P.~Gondolo, JCAP \textbf{04} (2007) 012.
\bibitem{lewin-smith} J.~D.~Lewin and P.~F.~Smith, Astropart. Phys. \textbf{6} (1996) 87.
\bibitem{n-body} J.~Diemand, B.~Moore and J.~Stadel, MNRAS {\bf 352} (2004) 535. [arXiv: astro-ph/0402160]
\bibitem{vel-dyn} A.~M.~Green, Phys.\ Rev.\ {\bf D66} (2002) 083003. [arXiv: hep-ph/0207366)
\bibitem{read} J.I.~Read \textit{et al.}, [arXiv: 0902.0009].
\bibitem{bruch} T.~Bruch \textit{et al.}, [arXiv: 0902.4001].
\bibitem{webtool} J.~Edsj\"{o}, WimpSim web tool, http://www.physto.se/\~{ }edsjo/wimpsim/scripts.html.
\bibitem{directrate} K.~Griest, Phys. Rev. D \textbf{38} (1988) 2357.
\bibitem{pythia} T. Sj\"ostrand \textit{et al.}, JHEP \textbf{05} (2006) 026.
\bibitem{cirelli} M.~Cirelli \textit{et al.}, Nucl.\ Phys.\ {\bf B727} (2005) 99; Erratum-ibib {\bf B790} (2008) 338. [arXiv: hep-ph/0506298]
\bibitem{angular_sel} M.~Mori \textit{et al.}, Phys. Rev. D \textbf{48} (1993) 5505.

\end{thebibliography}
\end{document}